\documentclass[twocolumn,epsf,prb,amsmath,amssymb,showpacs]{revtex4}
\usepackage{epsfig }
\usepackage{amsmath}
\usepackage{amssymb}

\DeclareMathAlphabet{\bi}{OML}{cmm}{b}{it}

\begin{document}
\title{ Wavevector filtering through single-layer and bilayer
graphene \\ with magnetic barrier structures }
\author{M. Ramezani Masir$^{1}$,  P. Vasilopoulos$^2$
and F. M. Peeters$^1$}
\address{
\ \\
$^{1}$Department of Physics, University of Antwerp,
Groenenborgerlaan 171, B-2020 Antwerpen.}
\address{
\ \\
$^{2}$Department of Physics, Concordia University, Montreal, Quebec,
Canada H3G 1M8}

\begin{abstract}
We show that the angular range of the transmission  through magnetic
barrier structures can be efficiently controlled in single-layer and
bilayer graphene and this renders the structures efficient
wavevector filters. As the number of magnetic barriers increases
this range shrinks, the gaps in the transmission versus energy
become wider, and the conductance oscillates with the Fermi energy.
\end{abstract}
 \maketitle

Graphene's electronic properties are drastically different from
those, say, of conventional semiconductors. Charge carriers in a
wide single-layer graphene
 behave like "relativistic", chiral massless particles  with a
"light speed" equal to the Fermi velocity and possess  a {\it
gapless,  linear} spectrum close to the $K$ and $K'$ points
\cite{novo1}. A  perfect transmission through arbitrarily high and
wide barriers, referred to as Klein tunneling \cite{kat} is
expected. In contrast, carriers in bilayer graphene possess a {\it
quadratic} spectrum  with {\it zero gap}. However, applying a gate
allows the opening of a gap due to the tunnel coupling between the
layers. This impedes the Klein tunneling and is  more appropriate
for certain applications, e. g., for improving the on/off ratio in
carbon-based transistors. A recent review of the properties of
graphene is given in Ref. \onlinecite{cast}.

An angular confinement of the transmission  in single-layer graphene
was reported in Refs. \onlinecite{egg} and \onlinecite{mass} using
magnetic barrier structures. However, in none of these works was the
wavevector filtering behavior of magnetic barriers addressed. In
this Letter we show how this angular confinement can be maximized,
in single-layer and bilayer graphene, and detail the resulting
filtering behavior. This occurs for  finite-width and
$\delta$-function barriers but not for complex magnetic structures
in which the average magnetic field is zero.

An electron in single-layer graphene, in the presence of a
perpendicular magnetic field $B(x)$, that  varies along the $x$
direction, is described by the $2\times 2$ Hamiltonian $
  H_0 = v_{F}\boldsymbol{\sigma}\cdot( \textbf{p} + e\textbf{A}(x))$.
A solution of  Dirac's equation $H_0\Psi=E\Psi$, in regions where
the magnetic field is not zero, is a linear combination of Weber
functions \cite{egg,mass}. For an electron in bilayer graphene the
relevant  Hamiltonian is a $4\times 4$ matrix \cite{milt, mass} with
elements $H_{13}=H_{31}=t$ and $t$ the tunnel coupling strength
assumed   constant  $t\approx 400meV$. Using the dimensionless units
$
 B(x) \rightarrow B_{0}B(x),
\,\,A(x)\rightarrow B_{0} \ell_{B} A(x), \,
    \vec{r} \rightarrow \ell_{B}\vec{r}, \,
    \vec{v} \rightarrow v_{F}\vec{v}, \, E \rightarrow E_{0} E , \, t \rightarrow E_{0} t$, with $ E_{0}=\hbar
v_{F}/\ell_{B}$ and $\ell_{B} =[\hbar/ eB_{0}]^{1/2}$, the energy
spectrum for a homogeneous magnetic field takes the form
\begin{equation}
\hspace*{-0.2cm}  E_{n,\pm} = \pm [2n + 1 + t^{2}/2 \pm
\sqrt{t^{4}/4 + (2n+1)t^{2} + 1} ]^{1/2}.
\end{equation}
For $t=0$ it gives the spectrum for single-layer graphene
\cite{milt}
$E=\pm[2(n+1)]^{1/2}$. The low-energy spectra for a  barrier on single-layer %%($t=0$)
and bilayer %%($t\neq 0$)
graphene are
%%%
partly
%%%
shown in the lower  panel of Fig. 1 as a function of $k_y$. For
single-layer graphene in a region of homogeneous magnetic field the
wave function, up to a normalization factor, is written in terms of
the Weber functions $D_p(z)$, $p = E^{2}/2$, as
\begin{equation}\label{wave2}
  \Phi(z)\sim \left(
  \begin{array}{c}
       -i (E/\sqrt{2})D_{p-1}(z)\\
       \,{D_{p}(z)}.
       \end{array}
     \right).
\end{equation}
\begin{figure}[ht]
\begin{center}
\includegraphics[width=5cm]{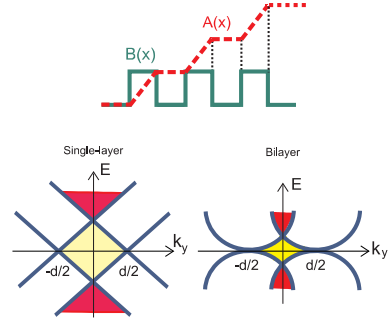}
\caption{Top panel: magnetic field profile $B(x)$ and corresponding
vector potential $A(x)$ for a series of barriers. Lower panel: a map
in $(E, k_y)$ space, showing the regions in which tunneling
involving real wave vectors is allowed (in red) and in which is not
(in yellow) for {\it one} magnetic barrier in single-layer and
bilayer graphene. Bound states can be found in the yellow regions
delimited by the free-particle spectra \cite{mass}.} \label{fig1}
\end{center}
\end{figure}
For bilayer graphene the corresponding wave function is

\begin{equation}\label{websol}
     \Phi^{\pm}(z) \sim \left(
       \begin{array}{c}
       D(p^{\pm}, z) \\
       \\
       (-i {\sqrt{2}p^{\pm}\over E})D(p^{\pm}-1, z) \\
       \\
       ({E\over t} - {2p^{\pm}\over t E})D(p^{\pm}, z)\\
       \\
       i {\sqrt{2}\over E} ({E\over t}-{2 p^{\pm}\over t E})D(p^{\pm}+1, z)
       \end{array}
     \right),
\end{equation}
where $p^{\pm} = (\gamma_{\pm} - 1)/2$, $\gamma_{\pm} = E^{2}\pm
(1+E^{2}t^{2})^{1/2}$, and $z=2^{1/2}(x+k_{y})$. In regions where
the magnetic field is zero the solution is written as ($T$ denotes
the transpose)
\begin{equation}\label{m8}
\hspace*{-0.2cm}
    \Psi^{R}_{E, \pm}=N_{\pm}
    \left(\mp E,\mp k_{x}^{\pm}\pm i k_{y},E, k_{x}^{\pm} + i
    k_{y}\right)^T
    e^{ik_{x}^{\pm}x + ik_{y} y},
\end{equation}
for  waves propagating to the right, and as
\begin{equation}\label{m8}
 \hspace*{-0.2cm}   \Psi^{L}_{E, \pm} = N_{\pm}\left(\mp E,\pm k_{x}^{\pm}\pm i
 k_{y}, E,
 -k_{x}^{\pm} + i k_{y}\right)^T e^{-ik_{x}^{\pm}x + ik_{y} y}
\end{equation}
for  waves propagating to the left.  In this case the energy
spectrum and wave vector $k_x^\pm$ are
\begin{eqnarray}
  E &=& \mp t/2 \pm
\left[t^{2}/4 + K^\pm\right]^{1/2},\\
 k^{\pm}_{x} &=& \left[E^2 - k_{y}^2 \pm E t\right]^{1/2},
\end{eqnarray}
where $K^\pm=k^{2\pm}_{x} + k_{y}^{2}$. The  factor $N_{\pm}=1/(4 W
E k_{x}^{\pm})^{1/2}$ can be obtained  from normalization of the
current.
To evaluate the transmission through
%%%
one or many barriers
%%%
we use the transfer-matrix technique and at each interface we match
the wave function and the flux \cite{mass}. Scattering between the K
and K' valleys is  negligible for fields below $10^4$ Tesla
\cite{park}. For a single barrier we can evaluate the angular
confinement by imposing the condition that the wave number $k_{x}$
be real for incident and transmitted waves. This gives
\begin{eqnarray}
k_{i} &=& [E^{2}- (k_{y}+A_{i})^{2}]^{1/2},\\
k^{\pm}_{i} &=& [(E \pm t/2)^{2} - t^{2}/4 - (k_{y} +
A_{i})^{2}]^{1/2},
\end{eqnarray}
 respectively, for a single layer and a bilayer.
Further, $i = 1,2$, $A_{1} = d/2$, and $A_{2} =- d/2$. From Eqs. (8)
and (9)  and the relations $k_{y} = E_{F}\sin{\theta}$  for a single
layer  and  $k_{y} = (E^{2}_{F} \pm E_{F}t)^{1/2}\sin{\theta^{\pm}}$
for a bilayer we can  find the range of the angle of incidence
$\theta$ in which the transmission is confined. For a single barrier
we obtain
\begin{equation}\label{hg14}
    -1\leq \sin{\theta} \leq 1 - d/E,
\end{equation}
in single-layer graphene, and
\begin{equation}\label{hg14}
    -1 \leq \sin{\theta^{\pm}} \leq 1 - d/(E^{2} \pm
    Et)^{1/2}.
\end{equation}
in bilayer graphene. In single-layer graphene the angle of exit
$\theta'$ is related to $\theta$ by
\begin{equation}\label{hg14}
    \sin\theta' = \sin\theta + d/E
\end{equation}
and in bilayer graphene by
\begin{equation}\label{hg14}
    \sin\theta'^\pm = \sin\theta^\pm + d/(E^2\pm Et)^{1/2}.
\end{equation}
We can also replace $\sin\theta$ by $k_y/k$ in Eqs. (10)-(13). If we
don't use the dimensionless units, the last terms in Eqs. (10)-(13)
are multiplied by $ \lambda=\hbar v_F/\ell_B^2$, i.e.,
\begin{equation}\label{hg14}
    d/E\to
    \lambda d/E, \,\, d/(E^2\pm Et)^{1/2}\to \lambda d/(E^2\pm Et)^{1/2}.
\end{equation}

From Eqs. (10) and (12) one can find the limits for $\theta'$.
Notice  that Eq. (10) implies a total reflection for $d>2E$. We
remark in passing that for the usual 2D Schr\"{o}dinger electrons
Eqs. (10) and (12) remain valid with $E$ replaced by $(2E)^{1/2}$.

By controlling the ratio $d/E$ or $d/ (E^{2}  \pm
    Et)^{1/2}$, Eqs. (10) and (11) show
how one can control the angular range of the transmission. For
instance, one could double the width of the barrier or raise $E_F$,
through the density, and/or $t$ by doping \cite{ota} or gating
\cite{gat}. However, one may not have a good, continuous control
over these parameters, e.g., over the width $d$, which may be
limited by fabrication techniques. It is therefore worth searching
for other ways to control these parameters.
\begin{figure}[ht]
\begin{center}
\includegraphics[width=7cm]{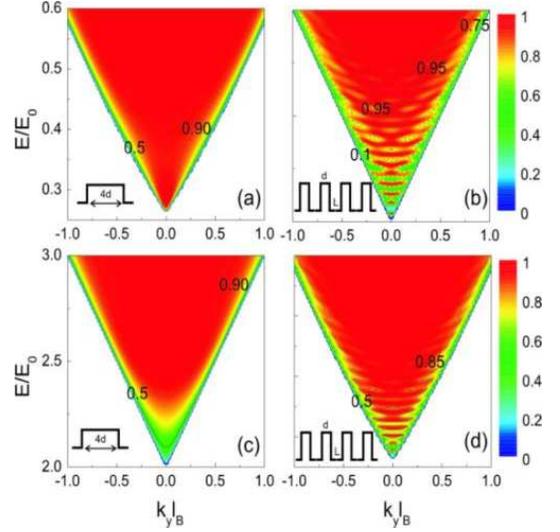}
\caption{Contour plot of the transmission through {\it one} magnetic
barrier of width $d=4\ell_B$ in (a) and through {\it four}  barriers
of width $d=\ell_B$ in (b) for single-layer graphene. (c), (d):
Respectively as in (a) and (d) for bilayer graphene with $t=15$.}
\label{fig2}
\end{center}
\end{figure}
With this in mind we investigate the effect of increasing the number
of barriers on the transmission. The  region  in which tunneling,
involving real wavevectors $k_x$, is allowed is shaded in red in
Fig. 1. For $n$ barriers  the angular range is given by Eqs. (10)
and (11) upon  changing $d$ to $nd$.
From Eqs. (10) and (11) we can evaluate the  minimum energy required
for tunneling. For $n$  barriers of width  $d$ and $\theta=0$ we
have $E = nd/2$ for single-layer graphene and $E = \mp (t/2) \pm
(t/2)[1+ n^{2} d^{2}/t^{2}]^{1/2}$ for bilayer graphene. For  $t\gg
nd$ we can rewrite this minimum as $E\approx \pm n^{2}d^{2}/4t$.
This shows clearly that we can control the  tunneling  by increasing
the number of barriers.
\begin{figure}[ht]
\begin{center}
\includegraphics[width=7cm]{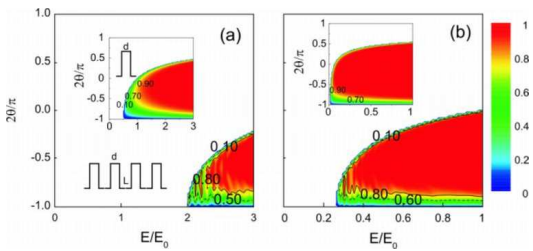}
\caption{(a) Contour plot of the transmission in ($E, \theta$) space
for {\it four} barriers of width $d=\ell_B$ with $L=10\ell_B$ in
bilayer graphene. The inset shows the result for {\it one} barrier
with $d=\ell_B$. (b) As in (a) but for single-layer graphene.}
\label{fig3}
\end{center}
\end{figure}
Typically the height of the magnetic barriers is in the range
0.1-1Tesla. For $B = 1 T$ the magnetic length is $ l_b= 25.6$ nm and
the energy scale $E_0= 25.7 $ meV for a typical $v_ F= 10^{6}$ m/s.
In Fig. 2 we show a contour plot of the transmission, in ($E, k_y)$
space, for {\it one}  barrier of width $d=4\ell_B$ and for {\it
four} magnetic barriers of width $d=\ell_B$.  As seen using {\it
four} barriers instead of {\it one} results in sharp resonances
shown in the lower part of the right panels for both single-layer
and bilayer graphene. This holds for any multibarrier structure
$n>2$.
\begin{figure}[ht]
\begin{center}
\includegraphics[width=7cm]{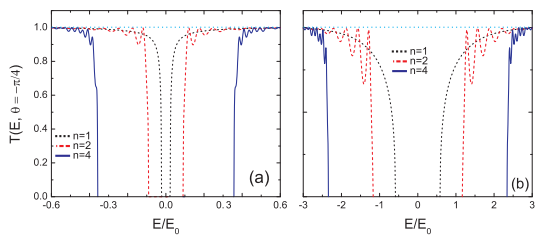}
\caption{Transmission probability  for different number of magnetic
barriers in bilayer graphene (a) and single-layer graphene (b) with
the parameters of Fig. 3.} \label{fig4}
\end{center}
\end{figure}
As Eq. (10) shows, the angular range of the transmission shrinks by
increasing the number of barriers. We show that in Fig. 3(a) for
bilayer graphene and in Fig. 3(b) for single-layer graphene. Had we
used {\it one} barrier of width $d=4\ell_b$, the shrinking of the
shown angular range would be the same but the resonaces at the lower
left end of the plot would be absent. This can be seen more clearly
in Fig. 4 where we plot the transmission versus the energy for
$\theta=-\pi/4$ and different number of barriers. Notice that the
energy gap widens with $n$.
Having seen the  transmission resonances, one may wonder to what
extent they survive the averaging over the Fermi surface that
determines the conductance $G$ which is more easily accessed in
experiments than the transmission. For  very low temperatures, such
that $f(E)-f(E+eV)\approx -eV\delta (E-E_F)$, with $f(E)$ the
Fermi-Dirac function and $eV$ the source-to-drain voltage drop, the
standard formula  for $G$ is
 $G = G_{0}\int^{\pi /2}_{-\pi /2}T(E_{F},E_{F}\sin{\theta}){\cos\theta}
d\theta$,  with $G_0=2 e^{2} E_{F} L_y / \pi h$ and $L_y$   the
length of the structure along the $y$ direction. In Fig. 5(a) we
plot the conductance of a four-barrier structure in single-layer
graphene and in Fig. 5(b) in bilayer graphene with $d=\ell_B$ and
two values of the interbarrier distance, $L = 10\ell_B$ and $L =
20\ell_B$, and $t=15$. The insets show the corresponding results for
a single barrier of width $d=4\ell_B$. As shown, in line with the
transmission results of Fig. 3 those for the conductance show
oscillations in the multibarrier structure that are absent in the
single barrier.
\begin{figure}[ht]
\begin{center}
\includegraphics[width=7cm]{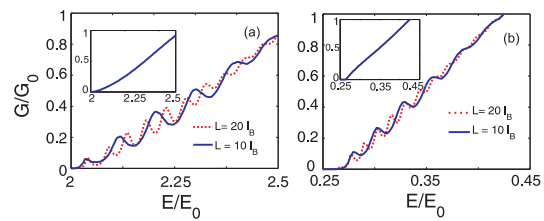}
\caption{ Conductance through {\it four} magnetic barriers, of width
$d=\ell_B$, in single-layer graphene (a) and in bilayer graphene (b)
for two values of $L$ and $t=15$. The upper insets show the result
through {\it one} magnetic barrier of width $d=4\ell_B$.}
\label{fig5}
\end{center}
\end{figure}
In summary, we showed that the angular range of  the transmission
through magnetic barrier structures can be efficiently controlled by
increasing the number of barriers in single-layer and bilayer
graphene.
This renders the structures efficient wavevector filters, leads to
transmission resonances  and conductance oscillations, and widens
the gaps in the transmission as a function of the energy.\\

This work was supported by the Flemish Science Foundation (FWO-Vl),
the Belgian Science Policy (IAP) and the Canadian NSERC Grant No.
OGP0121756.
 \vspace*{-0.5cm}

\end{document}